\documentclass[12pt,review,number]{elsarticle} 
\pdfoutput=1
\usepackage{graphicx}
\usepackage{natbib}

\journal{Computational Materials Science}
\begin{document}             % End of preamble and beginning of text.

\begin{frontmatter} 

\title{Improved Calculation of Vibrational Mode Lifetimes
      in Anharmonic Solids - Part II: Numerical Results}
\author{Doyl Dickel}
\author{Murray S. Daw}
\ead{daw@clemson.edu}
\cortext[cor1]{Corresponding author}
\address{Dept of Physics \& Astronomy, Clemson University, Clemson,
 SC 29634}
 
  \date{\today}

\begin{abstract}
  In a two-part publication, we propose and analyze a formal
  foundation for practical calculations of vibrational mode lifetimes
  in solids. The approach is based on a recursion method analysis of
  the Liouvillian. In the first part, we derived the lifetime of
  vibrational modes in terms of moments of the power spectrum of the
  Liouvillian as projected onto the relevant subspace of phase
  space. In practical terms, the moments are evaluated as ensemble
  averages of well-defined operators, meaning that the entire
  calculation is to be done with Monte Carlo. In this second part, we
  present a numerical analysis of a simple anharmonic model of
  lattice vibrations which exhibits two regimes of behavior, at low
  temperature and at high temperature. Our results show that, for this
  simple model, the mode lifetime as a function of temperature and
  wavevector can be simply approximated as a function of the shift in
  frequency from the harmonic limit. We next compare these
  calculations, obtained using both Monte Carlo and computationally
  intensive molecular dynamics, with those using the lowest order
  moment formalism from the Part I. We show that, in the
  high-temperature regime, the lowest order approximation gives a
  reliable approximation to the calculated lifetimes. The results also
  show that extension to at least fourth moment is required to obtain
  reliable results over a full range of temperatures.
  \end{abstract}
  

\end{frontmatter}

\section{Introduction}

The calculation of vibrational mode lifetimes using the Green-Kubo
relation\cite{kubo91} in solids is, at present, a computationally
expensive process. The formalism presented in Part I of this
work~\cite{daw09a} provides an approach to numeric computation of
these lifetimes utilizing only ensemble statistics, allowing for much
quicker computation. The strength of this formalism is greatly
enhanced as it does not require the use of computationally expensive
molecular dynamics needed to calculate autocorrelations of occupation
numbers in the solid up to large times. Rather, mode lifetimes can be
approximated by examining the properties of a given system over the
ensemble, which requires no molecular
dynamics (MD). \\

The mode lifetime is defined here by the Green-Kubo relation
\begin{equation}
\tau_k = \int_{-\infty}^{+\infty} dt\ \chi_k(t)
\label{eq:GKform}
\end{equation}
where $\tau_k$ is the lifetime of mode $k$, and $\chi_k(t)$ is the
mode autocorrelation function
\begin{equation}
\chi_k (t) = \frac{\langle \delta n_k(0)\ \delta n_k(t)
  \rangle}{\langle \delta n_k(0)^2 \rangle} 
\label{eq:chik}
\end{equation}
Here, $\delta n_k$ is the fluctuation of the occupation number of a
given mode, and the angular brackets indicate an average over the
equilibrium distribution.  The straightforward but tedious method of
calculating $\tau_k$ requires evolving a given state forward in time
using MD, calculating $\chi(t)$ and averaging over the equilibrium
ensemble. The improved method, introduced in Part 1, is based on
applying the recursion method\cite{haydock95,haydock99,haydock06} to
the Liouvillian operator\cite{koopman31,koopman32}, $\hat{L}$, defined
by
\begin{equation}
\hat{L} = i\{H, \} = i \sum_l (
\frac{\partial H}{\partial q_l} \frac{\partial}{\partial p_l} - 
\frac{\partial H}{\partial p_l} \frac{\partial}{\partial q_l} )
\label{eq:Liovillian}
\end{equation}
with $q$ and $p$ the coordinate and momentum variables of the system
and $H$ the Hamiltonian. By successive applications of the Liouvillian
on $\delta n_k$ we can generate a sequence of orthonormal
functions\cite{daw09a,haydock80:_recur_method}.  With this sequence as
a basis, the Liouvillian takes a special tridiagonal form. Using this
scheme, $\chi_k(t)$ can be simply related to the resolvent
$\hat{R}(\omega)$ of the Liouvillian (defined by $ \hat{R}(\omega) =
(\omega - \hat{L})^{-1} $). The resolvent can then be expressed as a
continued fraction. More precisely, the auto-correlation is related to
the projection of the resolvent onto $\delta n_k$. It can then be
shown that $\tau_k$ can be expressed in terms of the moments of
$\tilde{\chi}(\omega)$, the Fourier transform of $\chi(t)$. The power
of the method comes if it is possible to obtain a reasonable
approximation from only a few low-order
moments. \\

In order to evaluate the effectiveness of this method, we compare the
lowest-order calculation to numerically exact results for a simple
model of anharmonic lattice dynamics. To this end, we have determined
$\tau_k$ two ways, first, by calculating it directly using MD and
Monte Carlo, and second, using the recursion method described above,
truncated to lowest moment. We also characterize some of the
interesting behavior of the model used.  \\

\section{Methods}

Our model is based a continuous vector-like quantity defined on a
three-dimensional lattice. The underlying lattice structure is simple
cubic (8x8x8, with 512 sites) with periodic boundaries. Nearest
neighbor lattice sites are connected by anharmonic potentials such
that the Hamiltonian of the system is given by
\begin{equation}
H=\sum_i \frac{1}{2} \left|\vec{p}_i\right|^2 + \sum_{<i,j>} V(\vec{d}_i-\vec{d}_j)
\label{eq:H1D}
\end{equation}
where $\vec{p}_i$ is the momentum of the particle at lattice site,
$i$, and $V(\vec{d})$ is given by 
\begin{equation}
V(\vec{d})=\frac{1}{2} |\vec{d}|^2+\frac{1}{24} |\vec{d}|^4
\label{eq:potential}
\end{equation}
The coupling occurs for every pair $(i,j)$ which are
nearest-neighbors. \\

Note that this model describes a vector degree of freedom ($\vec{d}$)
on each lattice site, which is connected to neighboring degrees of
freedom. However, it does \emph{not} capture particle motion. In other
words, since the neighbors in the lattice are fixed, there is no mass 
flow, but only vibrations about a fixed lattice. In this model, there 
is no melting transition, which we consider to be an advantage because
we can examine the range of validity of our calculations over a 
virtually unlimited range of temperatures. This is in contrast to the
work of Ladd, \emph{et al.}\cite{ladd86} who studied mode lifetimes in a
Lennard-Jones system, which does have a melting transition, so that
their studies were effectively limited to a narrower range of
temperatures. Furthermore, by choosing a Hamiltonian which does not
include particle flow, we are able to selectively study the nature of
anharmonic \emph{lattice} vibrations. \\

We also note that this model displays different behavior in two
temperature regimes. At low temperature, the system behavior is
dominated by the harmonic part of the potential, with only a weak
anharmonicity evident in the dynamics. This is characterized by a
harmonic heat capacity and long lifetimes for the modes. At higher
temperature, the system is dominated by the quartic part of the
potential, which is characterized by a different heat capacity and
shorter lifetimes for the modes. The transition between the two
regimes appears to be at a temperature of
about 10. \\

In order to calculate $\chi_k(t)$, an initial position is chosen in
phase space, using a Monte Carlo sampling at some fixed temperture,
and this state is propagated forward in time using the velocity Verlet
algorithm. At regular time intervals, $\chi(t)$ can be calculated from
the current and initial states, following the canonical transformation
laid out in Part I to determine the occupation number, $n$. As is
noted there, we can minimize the fluctuation in $\chi(t)$ by carefully
choosing the frequency used in the transformation, effectively
choosing the anharmonic frequency, $\omega_{a}$ as our transformation
frequency. This can be done by choosing the frequency which maximizes
$\tau_{k}$, as is explained in Part I. $\chi(t)$ is then calculated as
an average over fifty-thousand initial Monte Carlo points. \\

In part I, we showed how $\tau(k)$ could be related to the moments
($\mu_n$) of the resolvent, suitably projected onto the appropriate
function in phase-space corresponding to the mode with wavevector
$k$. That is, we obtained the exact result:
\[ \tau(k) = F( \mu_2, \mu_4, \mu_6, ... ) \]
where the moment can be calculated directly from the Liouvillian by
\[ \mu_n = \frac{ \langle \delta n_k \hat{L}^n \delta n_k \rangle }{
  \langle ( \delta n_k )^2 \rangle } \]
This can be re-expressed using dimensional analysis\cite{barenblatt96} in the form
\[ \tau(k) = \tau_2(k) \tilde{F}( \gamma_4, \gamma_6, ... ) \]
where
\[\tau_2=\sqrt{\frac{1}{\mu_2}}\]
and the $\gamma_n$ are dimensionless forms of higher moments, such as
$ \gamma_4 = \mu_4/(\mu_2)^2 $. Roughly speaking, the second moment
describes the width of the appropriate part of the power spectrum
($\tilde{\chi}(\omega)$ is the Fourier Transform of $\chi(t)$), while
the $\gamma$'s describe the shape of that power spectrum. \\

The approximation that naturally suggests itself here is to assume
that whatever changes in the power spectrum occur with temperature,
they can be tracked by just a small number of the lowest moments of
the power spectrum (or equivalently, the second moment and a small
number of $\gamma$'s). The simplest form of this approxmation would be
to rely entirely on the second moment ($\mu_2$), which is equivalent
to assuming that the $\gamma_n$ are all independent of temperature
(that is, the power spectrum only changes width but not overall
shape). There has been considerable success in other types of physical
systems with this moment-approximation, most notably in the
calculation of electronic structure of
materials\cite{haydock80:_recur_method,FS,AC}. \\

We therefore evaluate here the second moment approximation, for which
$\tau(k) \propto \tau_2(k)$ where the proportionality constant is some
function of the $\gamma_n$'s, which are all assumed to be independent
of temperature. We apply two tests. First, we calculate $\tau(k)$
according to the Green-Kubo prescription (which involves MD), and the
$\tau_2(k)$ using Monte Carlo alone. If the second moment
approximation is successful, the ratio $\tau_k/\tau_2(k)$ should be
independent of wavevector and temperature. The second test is more
stringent. If the time-scale of mode decay is simply proportional to
$\tau_2$, then this should be reflected in the auto-correlations
$\chi(t)$. Namely, if we scale the \emph{time} variable in
auto-correlation function for each mode by $\tau_2$, we should see a
``data collapse''. In other words, plotting $\chi(t/\tau_2)$ for all
wavevectors and temperature should yield a universal curve. The
detailed shape of that universal, scaled auto-correlation is
determined by the shape parameters (the $\gamma_n$'s). The ``data
collapse'' test is therefore a more rigorous determination of the
validity of the second-moment approximation. \\

As we will show in the following, the high-temperature behavior is quite
well-represented by the second moment approximation, but that the
lower-temperature behavior shows significant deviation. This indicates
that more moments may be needed to accurately approximate the low 
temperature behavior. 

\begin{figure}
	\centering
		\includegraphics[width=1.00\textwidth]{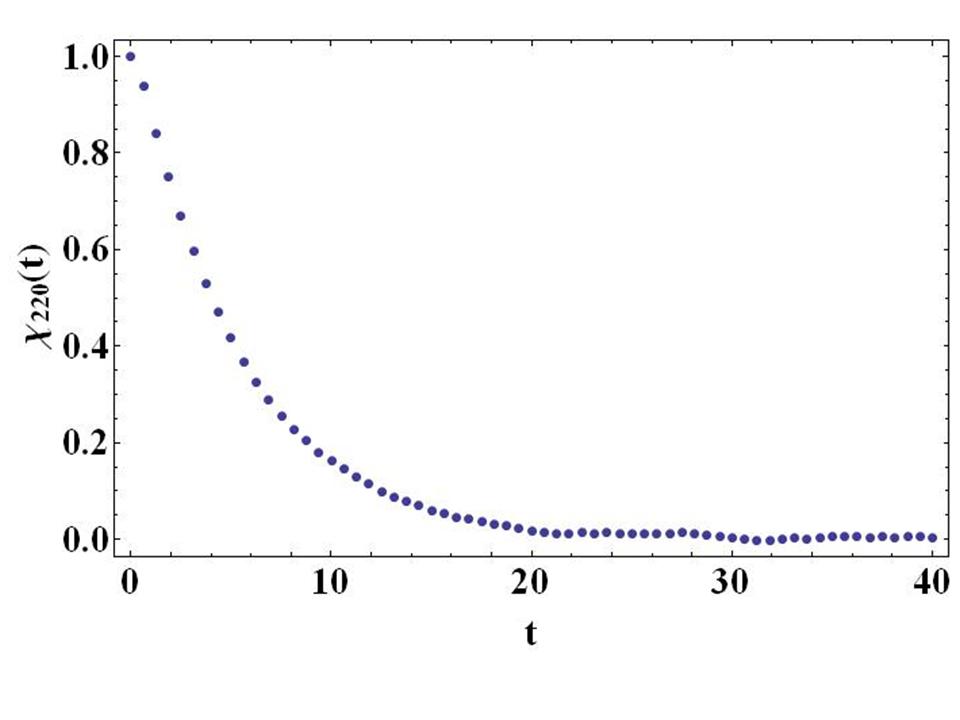}
	\caption{A typical autocorrelation, $\chi(t)$ for our model}
	\label{fig:Figure1_chit}
\end{figure}

\section{Numerical Results}

In this section, we review first the results of the numeric
calculation of the Green-Kubo lifetimes for our model Hamiltonian. A
sample of the autocorrelation as a function of time is given below 
(Fig.~\ref{fig:Figure1_chit}). It is first interesting to note that, due to the finite
size of the ensemble used, the autocorrelation does not go strictly
to zero for large times, so the termination of the autocorrelation is nontrivial. It would appear that, because of the finite number of
degrees of freedom, the ensemble retains some memory of its initial
state for arbitrarily large times. An examination of these residual
"tails'' as a function of system size shows that as the system becomes
large, the magnitude of the tails quickly goes to zero. For our finite
systems, the $\tau$ integral was truncated after the autocorrelation
dropped below 0.5 percent of the original value. \\

\begin{figure}
	\centering
		\includegraphics[width=1.00\textwidth]{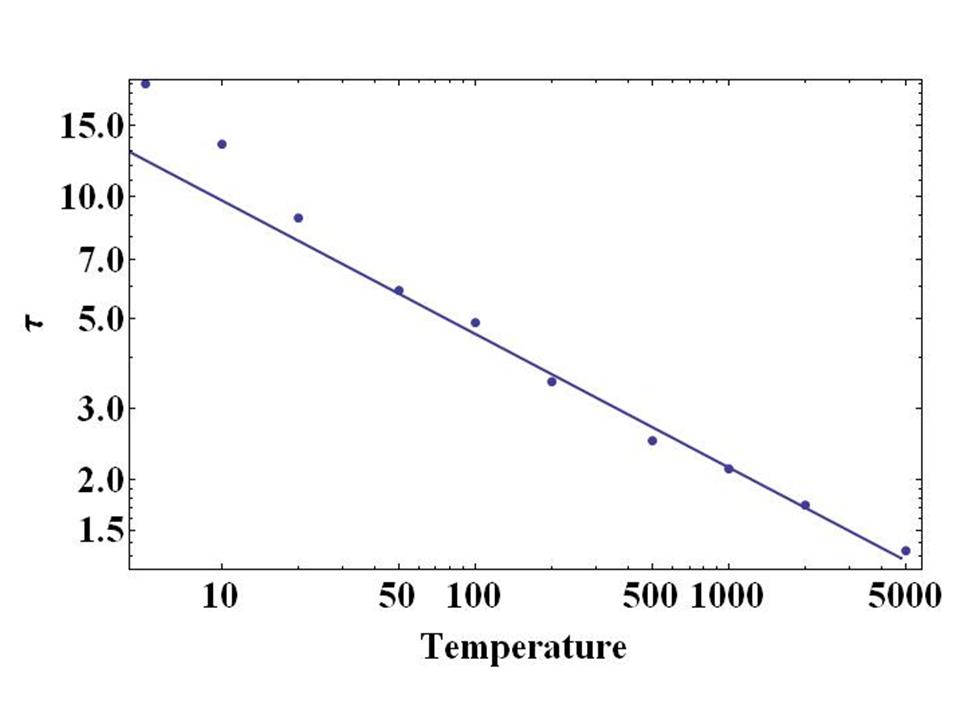}
			\caption{$\tau$ as a function of temperature
                          in the [220] direction. The solid line
                          represents $\tau \propto
                          T^{-1/3}$. Note that for temperatures less
                          than 10, the slope is steeper.} 
	\label{fig:Figure2_222}
\end{figure}

We first show how $\tau$ varies with temperature for a few wavevectors
and note that the general behavior is the same for all of them (Fig.~\ref{fig:Figure2_222}). As can be seen, the dependence on temperature to separate into two
regions, high and low temperature being above and below about 10. For
high temperatures, $\tau$ is related to temperature by a simple power
law, namely 

\begin{equation}
\tau(T)\propto T^{-\frac{1}{3}}
\label{eq:tauvT}
\end{equation}

This is different from the behavior discussed by Ladd for
Lennard-Jonesium, namely $\tau(T)\propto T^{-1}$. However, they expect
this behavior only at low temperature. We have not determined the
behavior of our model at low temperatures, because the computational
times increase substantially owing to the extended lifetimes at low
temperature. For the remainder of this paper, we will focus on the
high temperture regime and will only note apparent deviations as we
approach the transition to low temperature. \\

\begin{figure}
	\centering
		\includegraphics[width=1.0\textwidth]{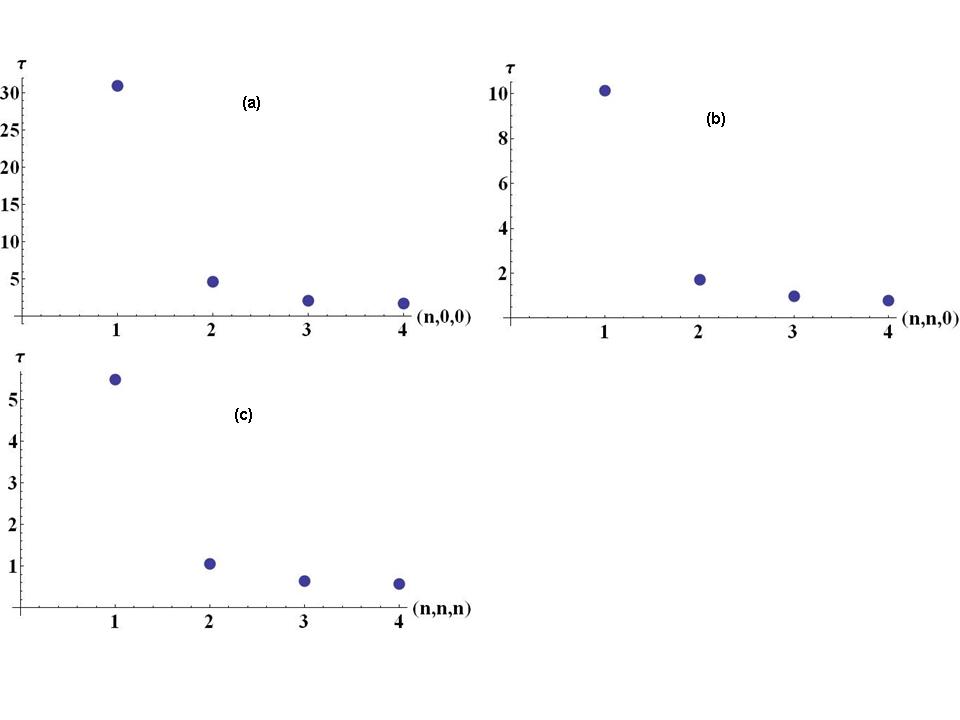}
	\caption{Tau as a function of wavevector in the (a) (1,0,0), (b) (1,1,0), and (c), (1,1,1) directions}
	\label{fig:Figure5_tauvk}
\end{figure}

Next we examine $\tau$ along the (100), (110), and (111)
directions (Fig.~\ref{fig:Figure5_tauvk}). We see here that, as expected, $\tau$ decreases with
increased wavevector (as well as increased frequency), but because of
the small size of the system, little more can be observed. \\

\begin{figure}
	\centering
		\includegraphics[width=1.00\textwidth]{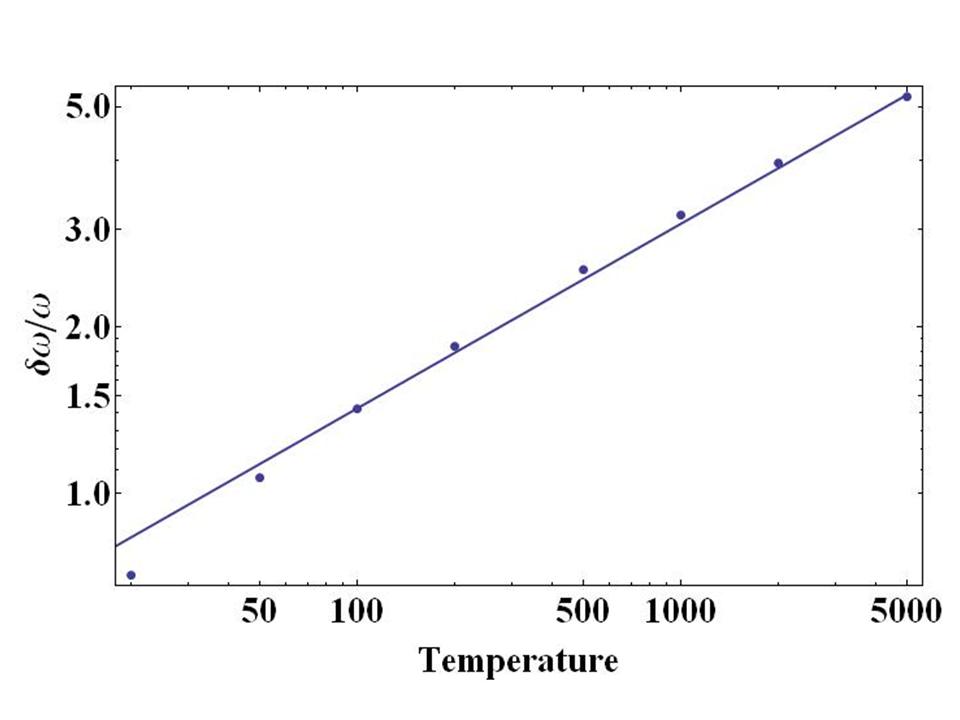}
                \caption{$\delta\omega/\omega$ as a function of
                  temperature. This ratio is the same regardless of
                  the wavevector examined. The solid line represents
                  $\delta\omega/\omega \propto T^{1/3}$}
	\label{fig:Figure3_domega}
\end{figure}

\begin{figure}
	\centering
		\includegraphics[width=1.00\textwidth]{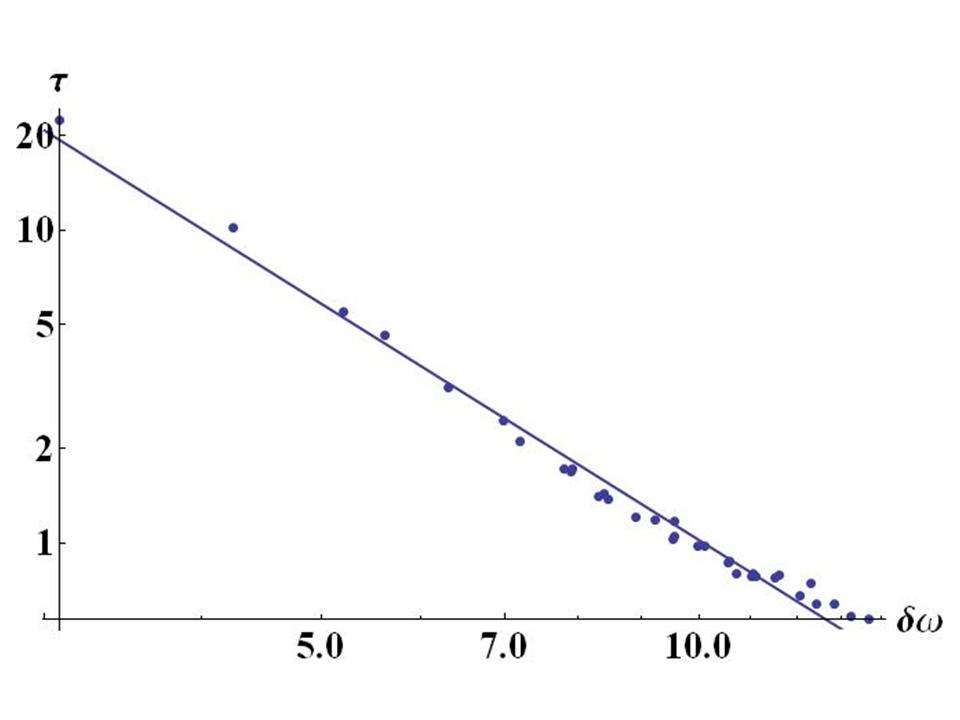}
	\caption{$\tau$ as a function of $\delta\omega$ at a temperature of 2000 for all nondegenerate wavevectors of the model used.}
	\label{fig:Figure4_tauvdomega}
\end{figure}

In order to see more directly how $\tau$ is determined by the
temperature and wavevector for this Hamiltonian, it is instructive to
examine the behavior of the anharmonic mode frequency
($\omega_a(k)$). For our model Hamiltonian, we find a simple relation
between the anharmonic frequency, the temperature, and the frequency in the
harmonic limit ($\omega_0(k)$). The harmonic frequency is of the mode
with wavevector $\vec{k}$ is \\
\begin{equation}
\omega_0(\vec{k})=\sqrt{6-2\cos{\frac{2\pi k_x}{l_x}}-2\cos{\frac{2\pi k_y}{l_y}}-2\cos{\frac{2\pi k_z}{l_z}}}
\label{eq:homega}
\end{equation}
As can be seen in Fig.~\ref{fig:Figure3_domega}, the ratio between the frequency shift ($\delta \omega(\vec{k}) \equiv \omega_a(\vec{k}) - \omega_0(\vec{k}) $)
and $\omega_0$, depends only on temperature (that is, \emph{not on wavevector}), and is
well approximated by the power law 
\begin{equation}
\frac{\delta\omega (\vec{k})}{\omega_0 (\vec{k})}=f(T)=(T/T_0)^{\frac{1}{3}}
\label{eq:domega}
\end{equation}
While the exponent and possibly the simple form of the relationship is
specific to our Hamiltonian, it is still an intriguing result. Based
on the similarities in the exponent in Eq. ~\ref{eq:tauvT} and ~\ref{eq:domega}, one reasonably suspects a relationship between the $\delta\omega$ and $\tau$. This
suspicion is confirmed by examining $\tau$ as a function of $\delta
\omega$ (see Fig.~\ref{fig:Figure4_tauvdomega}) for all wavevectors at a constant temperature,
where we see again a power-law dependence.   \\

The implication of this relationship is that, at a fixed temperature
the lifetime of a given mode depends dominantly on the anharmonic
frequency. This frequency, in turn, depends on two independent
factors, the harmonic frequency and the temperature. \\

Knowing the mode lifetimes as a function of wavevector and temperature
through our numerical simulation, we can now turn to the approximate
methods discussed in the companion work. \\

\section{Approximate Methods}

Because its determination does not require the use of any molecular
dynamics, $\tau_2$ can be calculated with \emph{much} less
computational expense then the numerical approach considered
above. Generally we have found that 200,000 MC steps is sufficient to
converge $\tau_2$ for all temperatures and wavevectors considered
here. \\

If we first examine the ratio $\tau_k/\tau_2$ as a function of
temperature for a particular wavevector, we see that, in the high
temperature regime, the ratio quickly converges to constant value
(Fig.~\ref{fig:Figure6_tauotaucollapse}). While only one wavevector is
shown, note that this behavior holds for all wavevectors. This is
expected, due to the separable nature of the anharmonic frequency for
our Hamiltonian and the dependance of $\tau$ on this frequency
mentioned above. Because $\tau$ can be determined from the shift in
frequency, the rest of them should display the same behavior. Because
the individual wavevectors may have different shapes, the ratio may
converge to different values. However, it remains true that this value
is constant for a given wavevector for sufficiently high
temperatures. The implication is that only $\tau_2$ is changing with
temperature, and the $\gamma$ factors discussed above
are constant. \\

\begin{figure}
	\centering
		\includegraphics[width=1.00\textwidth]{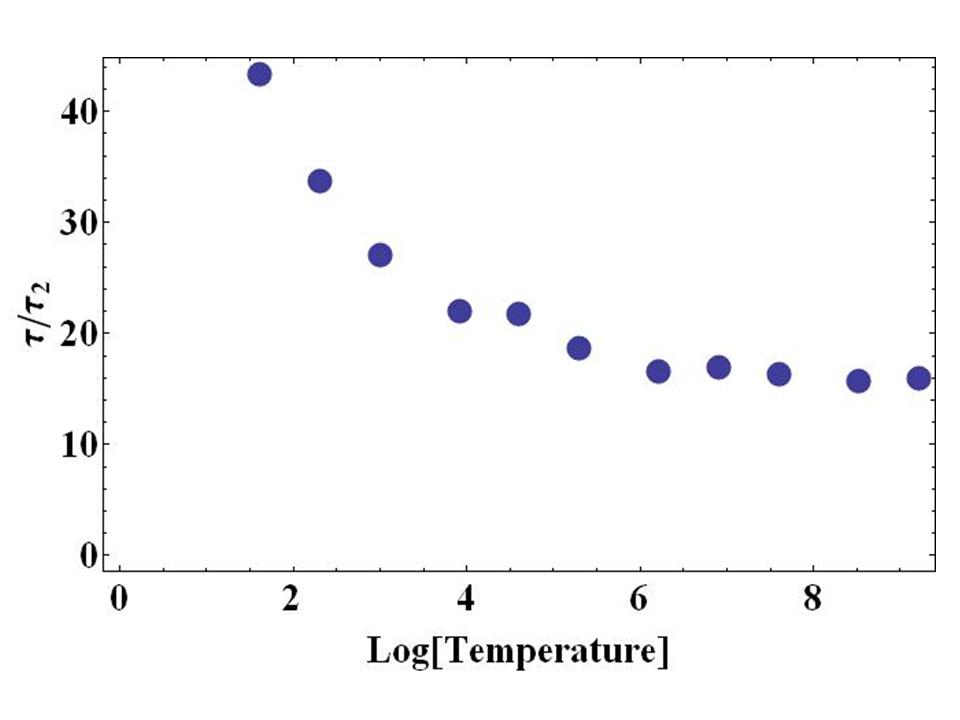}
	\caption{$\tau/\tau_2$ as a function of
                          temperature. We see that $\tau_2$ is
                          sufficient determine $\tau$ for high enough
                          temperatures.}
	\label{fig:Figure6b_tauotau}
\end{figure}

\begin{figure}
	\centering
		\includegraphics[width=1.0\textwidth]{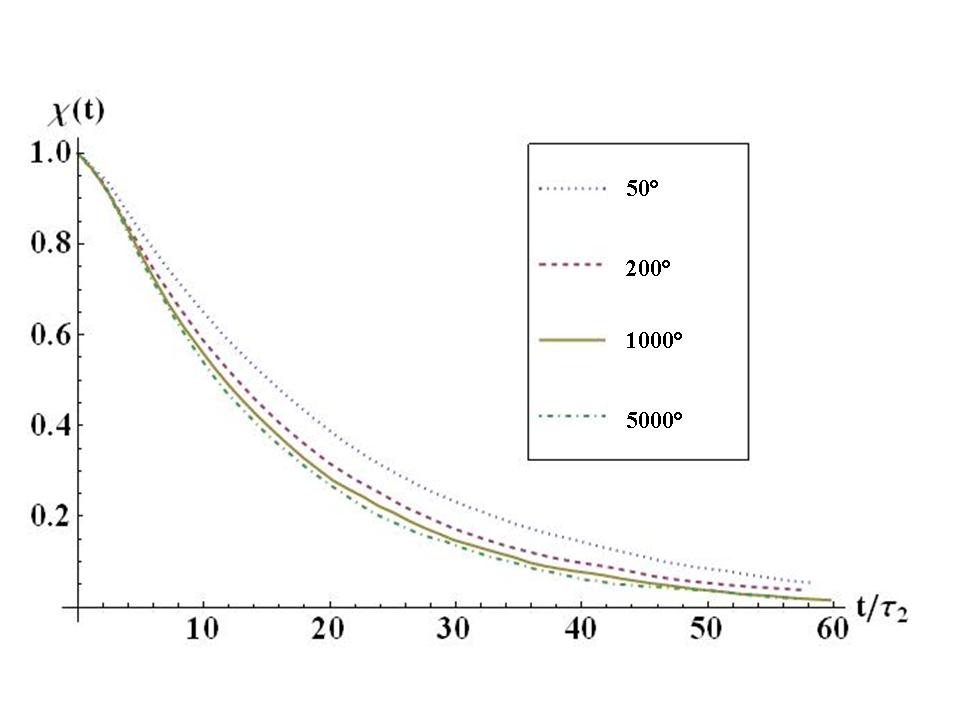}
                \caption{The data collapse of $\chi(t)$ for the [222]
                  wavevector at several different
                  temperatures. Because the data approach a universal
                  curve, $\chi(t)$ and therefore $\tau$ are
                  well-approximated, up to a constant, only by $\tau_2$.}
	\label{fig:Figure6_tauotaucollapse}
\end{figure}

With this notion that the shape (as opposed to the width) of $\chi(t)$
is invariant, we also test whether rescaling the elapsed time with
respect to $\tau_2$ will cause the data to collapse onto the same
curve. This is a more stringent test than the previous one as it
indicates that $\chi(t)$ and not just $\tau$ can be rescaled with
$\tau_2$. Note, in fact, that the collapse of $\chi(t)$ implies the
convergence of $\tau_k/\tau_2$ seen in
Fig.~\ref{fig:Figure6b_tauotau}. As
Fig.~\ref{fig:Figure6_tauotaucollapse} shows, this is clearly the case in 
the high temperature regime. In this limit, approximating $\tau$ at the
level of the second moment is sufficient to determine its behavior as
a function of temperature. \\

Because this agreement begins to break down at lower temperatures, it
may be neccesary to include higher moments as the shape of $\chi(t)$
changes from the high temperature limit. As the shape of the
autocorrelation changes, some number of the $\gamma$ factors are also
changing. By calculating them, and taking into account their effect on
the system, a more accurate prediction of $\tau$ should be possible. \\

\section{Conclusions}

By analyzing the mode lifetimes, through $\chi(t)$, in an anharmonic
lattice, with a simple, quartic interaction between nearest neighbors,
in three dimensions, we can draw several important conclusions. First,
for this particular Hamiltonian, the lifetime of a given mode depends
in a simple way on the anharmonic frequency of the mode, which in turn
depends separably on the harmonic frequency and the temperature. Both
of these relationships would seem to be peculiar to our model
Hamiltonian, but they allow a simplified analaysis of the
results. Second, seen as a function of temperature and frequency
directly, $\tau$ shows a simple behavior. More generally, and most
central to this work, the methodology developed in Part I is shown to
be a reliable convergent method of determining the mode lifetimes with
much lower computational cost than any present method of which the
authors are aware.  While deviations are observed for low
temperatures, related to the form of the potential used, using only
the first non-zero moment, $\chi(t)$ collapses effectively onto a
universal curve, and $\tau$ is correctly approximated at high
temperatures up to a constant to within a few percent. \\

While the Hamiltonian used here is a relatively simple one, which might be analyzed effectively using more direct methods, the scope of this new method should not be understated. Since calculation of lifetimes using moments requires no molecular dynamics, only averages over the ensemble, the computer time required for a calcuation is reduced dramatically, in the authors' experience, by at least an order of magnitude or more at high temperatures, where $/tau$ is small and even more for lower temperatures as $/tau$ increases and molecular dynamics calculations must go to higher times. Furthermore, the method can be readily extended to more complicated systems. As long as $\delta n$ can be found for a member of the ensemble, $\tau_2$ can be derived. In addition, this formalism provides a new way of examining vibrational mode lifetimes which may provide new insights into their behavior. This improved method offers not only significant advantages in both improving the methodology of calculation but also possible insights into the mechanism of dissipation. \\

\section{Acknowledgements}

This work was partly supported by DOE (\#DE-FG02-04ER-46139) and South
Carolina EPSCoR.

\newpage
\bibliographystyle{unsrt}
\bibliography{ILTC09_v3}

\end{document}